\documentclass[runningheads]{llncs}
\usepackage[backend=biber,style=authoryear,giveninits=true,uniquename=init]{biblatex}
\bibliography{biblio.bib}

\usepackage{amsmath}
\usepackage{booktabs} % For pretty tables
\usepackage{graphicx}
\usepackage{pgfplots}
\usepackage[all]{nowidow}
\usepackage[utf8]{inputenc}
\usepackage{tikz}
\usetikzlibrary{er,positioning,bayesnet}
\usepackage{multicol}
\usepackage{algpseudocode,algorithm,algorithmicx}
\usepackage{xfrac}
\usepackage{hyperref}
\hypersetup{
    colorlinks=true,
    linkcolor=blue,
    filecolor=blue,      
    urlcolor=blue,
    citecolor=black
    }
\usepackage[inline]{enumitem} % Horizontal lists

\usepackage[LGR,T1]{fontenc}
\newcommand{\textgreek}[1]{\begingroup\fontencoding{LGR}\selectfont#1\endgroup}

\makeatletter

\patchcmd{\NAT@test}{\else \NAT@nm}{\else \NAT@nmfmt{\NAT@nm}}{}{}

\DeclareRobustCommand\parenciteos
  {\begingroup
   \let\NAT@nmfmt\NAT@posfmt% ...except with a different name format
   \NAT@swafalse\let\NAT@ctype\z@\NAT@partrue
   \@ifstar{\NAT@fulltrue\NAT@citetp}{\NAT@fullfalse\NAT@citetp}}

\let\NAT@orig@nmfmt\NAT@nmfmt
\def\NAT@posfmt#1{\NAT@orig@nmfmt{#1's}}

\makeatother

\definecolor{blue}{HTML}{1F77B4}
\definecolor{orange}{HTML}{FF7F0E}
\definecolor{green}{HTML}{2CA02C}

\pgfplotsset{compat=1.14}

\begin{document}
\title{Smooth Pycnophylactic Interpolation Produced by Density-Equalising Map Projections}

\titlerunning{Smooth Pycnophylactic Interpolation}
\author{Michael T. Gastner$^\ast$, Nihal Z. Miaji and Adi Singhania}
\authorrunning{Gastner, Miaji and Singhania}
\institute{Yale-NUS College, 16 College Avenue West, \#01-220, Singapore 138527\\
$^\ast$Corresponding author: \email{michael.gastner@yale-nus.edu.sg}}
\maketitle              % typeset the header of the contribution
\begin{abstract}
A large amount of quantitative geospatial data are collected and aggregated in discrete enumeration units (e.g.\ countries or states). Smooth pycnophylactic interpolation aims to find a smooth, nonnegative function such that the area integral over each enumeration unit is equal to the aggregated data. Conventionally, smooth pycnophylactic interpolation is achieved by a cellular automaton algorithm that converts a piecewise constant function into an approximately smooth function defined on a grid of coordinates on an equal-area map. An alternative approach, proposed by Tobler in 1976, is to construct a density-equalising map projection in which areas of enumeration units are proportional to the aggregated data. A pycnophylactic interpolation can be obtained from the Jacobian of this projection. Here, we describe a software implementation of this method. Although solutions are not necessarily optimal in terms of predefined quantitative measures of smoothness, our method is computationally efficient and can potentially be used in tandem with other methods to accelerate convergence towards an optimal solution.

\keywords{spatial interpolation \and pycnophylactic density \and contiguous cartogram \and flow-based algorithm.}

\end{abstract}
\section{Introduction}

Quantitative geospatial data are often available only as aggregated numbers for discrete enumeration units. For example, national statistics agencies usually report the number of individuals living in each administrative division of a country (e.g.\ a census block in the United States and an Output Area in the United Kingdom) but do not release information about each individual’s exact location. Because it is impossible to infer exact locations, the aggregated data are often converted into a density function (in units of people per square kilometre) that assigns a real-valued number to each point belonging to the continuum of coordinates in the country. Let us assume that a country is divided into enumeration units \(U_1,\ldots, U_n\). We denote the population count in \(U_i\) by \(P_i\) and the area by \(A_i\). Furthermore, we assume that coordinates have already been converted from longitude and latitude to Cartesian coordinates \((x, y)\) using an equal-area map projection. A bounded, nonnegative function \(\rho(x, y)\) is referred to as a pycnophylactic density (from the Greek words \textgreek{πυκνός}, meaning ‘dense’, and \textgreek{φύλαξ}, meaning ‘guard’) if the aggregate density in \(U_i\) is equal to the observed population \(P_i\):

\begin{equation}
\iint_{U_{i}} \rho(x, y)\, d x\, d y=P_{i} \quad \text { for } i=1, \ldots, n
\end{equation}

The variable \(P_i\) does not need to be population in a narrow sense; it may also refer to other data that are only available in aggregated form (e.g.\ gross regional product or CO$_2$ emissions by enumeration unit). Henceforth, we use the term ‘population’ for any type of nonnegative, spatially extensive aggregated data.

An example of a pycnophylactic density is the following piecewise constant function:
\begin{equation}
\rho_\text{plateau}(x, y)=\left\{\begin{array}{cl}
P_{i} / A_{i}-\bar{\rho} & \text { if }(x, y) \in U_{i}, \\
0 & \text { if }(x, y) \notin \bigcup_{i} U_{i},
\end{array}\right.
\end{equation}
where
\begin{equation}
\bar{\rho}=\frac{\sum_{i} P_{i}}{\sum_{i} A_{i}}
\end{equation}
is the spatially averaged density. In principle, the density outside all enumeration units can be chosen arbitrarily because it cannot be inferred from the available data. However, it turns out to be mathematically convenient to impose the condition
\begin{equation}
\forall(x, y):(x, y) \notin \bigcup_{i} U_{i} \rightarrow \rho(x, y)=\bar{\rho},
\end{equation}
which can be statistically interpreted as imputation of missing data by substituting the mean. Henceforth, we apply equation (4) as a condition on any pycnophylactic density. Figure 1(a) illustrates the definition of \(\rho_\text{plateau}\), using COVID-19 cases in Croatia between 25 February 2020 and 24 March 2022 \parencite{croatian_institute_of_public_health_sluzbena_2022}. In this example, the density is obtained by dividing the number of COVID-19 cases in a county by the county’s area. Although \(\rho_\text{plateau}\) is easily calculated, piecewise constant densities are unsuitable statistical models for COVID-19 cases and many other geospatial data because of the discontinuities at the boundaries of enumeration units. As pointed out by \textcite{openshaw_ecological_1984}: ‘The areal units used to report census data (enumeration districts, census tracts, wards, local government units) have no natural or meaningful geographical identity.’ Thus, to reduce the artefacts introduced by arbitrary boundaries of enumeration units, it is generally preferable to work with a smooth  density, as illustrated in Figure 1(b), instead of \(\rho_\text{plateau}\), whose shape strongly depends on the location of the boundaries.

\begin{figure}
\includegraphics[width=\textwidth,height=\textheight,keepaspectratio]{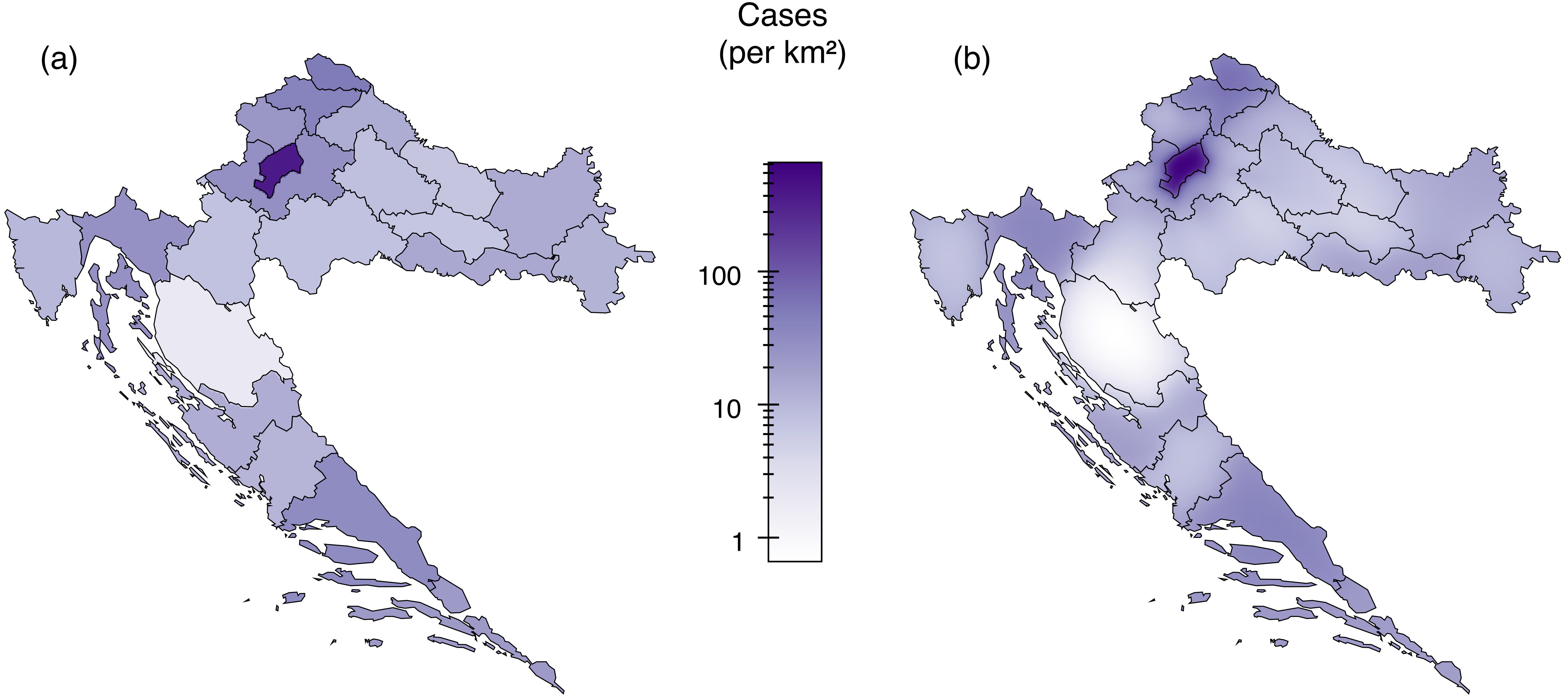}
\caption{Two pycnophylactic densities that represent the distribution of COVID-19 cases in Croatia between 25 February 2020 and 24 March 2022. (a) Piecewise constant density. (b) Smooth density obtained from a density-equalising map projection.}
\end{figure}

\textcite{tobler_smooth_1979} introduced a cellular automaton algorithm for generating smooth pycnophylactic densities, which approximates the continuum of space with a fine-grained square grid. Each point \((x, y)\) on the grid is initially assigned the density \(\rho_\text{plateau}(x, y)\). Thereafter, the density associated with each grid point is adjusted so that the absolute value of the discrete Laplacian is reduced, subject to the constraints that the sum of densities in each enumeration unit is conserved and density remains nonnegative. This adjustment is iterated until the changes are below a small threshold value or until the number of iterations reaches a predefined limit.

\citeauthor{tobler_smooth_1979}'s (\citeyear{tobler_smooth_1979}b) algorithm has become the standard technique for smooth pycnophylactic interpolation. The algorithm has been implemented in R \parencite{brunsdon_pycno_2014} and Python \parencite{noauthor_pysal_2021}. For users of Geographic Information Systems software, the algorithm is also available via extensions of ArcGIS \parencite{qiu_development_2012} and GRASS \parencite{metz_vsurfmass_nodate}. A variant of \citeauthor{tobler_smooth_1979}'s (\citeyear{tobler_smooth_1979}b) algorithm was developed by \textcite{rase_volume-preserving_2001}, in which the regular square grid is replaced by an irregular triangular network. However, the algorithm by \textcite{rase_volume-preserving_2001} and its later refinement by the same author \parencite{rase_volumenerhaltende_2007} keep the essential features of \citeauthor{tobler_smooth_1979}'s (\citeyear{tobler_smooth_1979}b) algorithm: iterative local averaging and subsequent redistribution of density differences to enforce the pycnophylactic condition of equation (1). Despite the widespread use of \citeauthor{tobler_smooth_1979}'s (\citeyear{tobler_smooth_1979}b) algorithm, it was not the first method proposed by him for smooth pycnophylactic interpolation. In an earlier publication, \textcite{tobler_cartograms_1976} described an alternative approach in which the boundaries of enumeration units are transformed into an area cartogram (i.e.\ a map in which all enumeration units are depicted with an area proportional to their population). \citeauthor{tobler_cartograms_1976}'s (\citeyear{tobler_cartograms_1976}) proposed method requires the area cartogram to be contiguous; that is, neighbouring enumeration units on the surface of the earth must be neighbours in the cartogram. As noted by \textcite{lapaine_choosing_2017}, contiguous cartograms are closely related to density-equalising map projections. In this study, we briefly review the connections between contiguous cartograms, density-equalising map projections and smooth pycnophylactic interpolation. Thereafter, we explain how to achieve smooth pycnophylactic interpolation using a recently developed algorithm that generates contiguous cartograms.

\section{Relationship between pycnophylactic interpolation and density-equalising map projections}

To construct a contiguous cartogram, the boundaries of enumeration units are modelled as polylines. We denote the vertices of the polylines on an equal-area map projection by \((v_{1x},v_{1y}), (v_{2x},v_{2y}),\ldots\) The vertices are then shifted to new positions \((w_{1x},w_{1y}), (w_{2x},w_{2y}),\ldots\) such that the regions demarcated by the transformed boundaries have an area proportional to the population of the corresponding enumeration units. This transformation can be regarded as the result of a map projection \(\mathbf{t} = (t_x, t_y)\) that satisfies the conditions \({t}_x\left(v_{jx},v_{jy}\right)=w_{jx}\) and \({t}_y\left(v_{jx},v_{jy}\right)=w_{jy}\) for all \(j=1,2,\ldots\) A map projection with this property can be obtained, for example, by solving the following equation:
\begin{equation}
\operatorname{det} J_\mathbf{t}(x, y)=\frac{\rho(x, y)}{\bar{\rho}}\ ,
\end{equation}
where \(\det{J_{\mathbf{t}}}=({\partial t_x}/{\partial x})({\partial t_y}/{\partial y})-({\partial t_y}/{\partial x})({\partial t_x}/{\partial y})\) is the Jacobian determinant of \(\mathbf{t}\), \(\rho(x, y)\) is a pycnophylactic density [i.e.\ it satisfies equation (1)] and \(\bar{\rho}\) is its spatial average given by equation (3). The quantity
\begin{equation}
\rho_{\text{res}}(x, y)=\rho(x, y)-\bar{\rho}\, \operatorname{det} J_{\mathbf{t}}(x, y)
\end{equation}
can then be interpreted as ‘residual density’ (i.e.\ the difference from the mean density that remains unexplained by the projection \(\mathbf{t}\)), and the objective is to find a solution \(\mathbf{t}\) such that \(\rho_\text{res}(x, y) = 0\) for all \((x, y)\).

Contiguous cartograms and density-equalising map projections are related to each other in two ways. First, if \(\rho\) and \((v_{jx}, v_{jy})\) are known, it is possible to solve equation (5) to obtain \(\mathbf{t}(x, y)\) and then obtain the polyline vertices \(\left(w_{jx},w_{jy}\right)\) of a contiguous cartogram by applying \(\mathbf{t}(x, y)\) to \(\left(v_{jx},v_{jy}\right)\) for all \(j=1,2,\ldots\) Second, if \(\left(v_{jx},v_{jy}\right)\) and \(\left(w_{jx},w_{jy}\right)\) are known, it is possible to construct a density-equalising map projection \(\mathbf{t}(x, y)\) with the property \(\mathbf{t}\left(v_{jx},v_{jy}\right)=\left(w_{jx},w_{jy}\right)\) and then obtain a pycnophylactic density by inverting equation (5):

\begin{equation}
\rho(x, y)=\bar{\rho}\, \operatorname{det} J_{\mathbf{t}}(x, y).
\end{equation}

In both cases, solutions are not unique. If \(\rho\left(x,y\right)\) is given, equation (5) allows infinitely many solutions \(\mathbf{t}(x, y)\) because the number of constraints implicit in equation (5) is 1, which is less than the number of dimensions (i.e.\ 2) of a geographic map. Conversely, if all \(\left(w_{jx},w_{jy}\right)\) are given, there are infinitely many ways to extend a function \(\mathbf{t}\) with the property \(\mathbf{t}\left(v_{jx},\ v_{jy}\right)=\left(w_{jx},w_{jy}\right)\) to the entire mapping domain. One could impose additional constraints on \(\mathbf{t}(x, y)\) to make the solution unique. At first glance, an obvious constraint would be to demand that \(\mathbf{t}(x, y)\) be conformal. However, conformality adds two constraints to the problem in the form of the Cauchy–Riemann equations; hence, together with the constraint of satisfying equation (5), the problem of finding \(\mathbf{t}(x, y)\) would be overdetermined \parencite{gastner_diffusion-based_2004}. Instead of demanding strict conformality, \textcite{tobler_continuous_1973} proposed that deviations from conformality should, at least, be minimised when constructing cartograms. However, he reported that a computer program, designed to find nearly conformal density-equalising map projections, failed to converge. Furthermore, it is not evident that a conformal map projection \(\mathbf{t}(x, y)\) necessarily generates desirable properties for a pycnophylactic density \(\rho(x, y)\), calculated using equation (7). Therefore, we describe a method that is guaranteed to find a density-equalising map projection \(\mathbf{t}(x, y)\) and a smooth pycnophylactic density \(\rho(x, y)\), even if neither \(\mathbf{t}(x, y)\) nor \(\rho(x, y)\) satisfy predefined criteria of optimality.

\section{Obtaining a pycnophylactic interpolation from a density-equalising map projection}

\textcite{gastner_pnas} introduced a flow-based algorithm that generates density-equalising map projections. In this algorithm, \(\rho(x, y)\) in equation (5) is treated as the initial density of a fluid. The algorithm proceeds by constructing a velocity field that conserves the mass of the fluid, is free of vortices and equilibrates the density over time. By integrating the velocity, the algorithm determines the final displacement \(\mathbf{t}(x, y)\) of any arbitrary point that is initially at position \((x, y)\). It can be shown that \(\mathbf{t}(x, y)\) satisfies equation (5); thus, \(\mathbf{t}(x, y)\) is a density-equalising map projection.

For the boundary conditions chosen by \textcite{gastner_pnas}, \(\mathbf{t}(x, y)\) can be calculated efficiently by applying a Fourier transform to the residual density \(\rho_\text{res}(x, y)\). Suppose that \(\rho(x, y)\) is initially chosen  to be the piecewise constant function \(\rho_\text{plateau}\) given by equation (2) on the unprojected map, where \(\det{J_{\mathbf{t}}}=1 \ \forall(x, y)\). Hence, the initial residual density can be expressed as follows:
\begin{equation}
\rho_{\text{res}, 1}(x, y)=\begin{cases}
P_{i} / A_{i}-\bar{\rho} & \text { if }(x, y) \in U_{i} \\
0 & \text { if }(x, y) \notin \bigcup_{i} U_{i}
\end{cases},
\end{equation}
where \(\bar{\rho}\) is the spatial average, defined in equation (3). Figure 2(a) shows \(\rho_\text{res,1}(x, y)\), using COVID-19 cases in Croatia as an example. Because \(\rho_\text{res,1}(x, y)\) has discontinuities at the boundaries of enumeration units, the Fourier transform of \(\rho_\text{res,1}(x, y)\) exhibits the Gibbs phenomenon \parencite{carslaw_gibbs_1925}; that is, the approximation of \(\rho_\text{res,1}(x, y)\) by a finite Fourier series exhibits large oscillations at the boundaries. These oscillations can cause numerical artefacts in subsequent calculations. To circumvent this problem, \citeauthor{noauthor_mgastnercartogram-cpp_nodate}'s (2022) computer implementation of the flow-based algorithm removes discontinuities from \(\rho_\text{res,1}(x, y)\) by applying Gaussian smoothing as a low-pass filter. Figure 2(b) illustrates the effect of Gaussian smoothing. The density after Gaussian smoothing can be calculated rapidly in Fourier space. However, Gaussian smoothing tends to shift density towards sparsely populated regions; thus, it does not produce a pycnophylactic density. Consequently, the calculated map projection \({\mathbf{t}}_1(x, y)\) is not density-equalising. Nevertheless, by projecting the boundaries, densely populated enumeration units tend to expand, and sparsely populated enumeration units tend to shrink, as shown by the county borders in Figure 2(c). We note that the algorithm replenishes the density that leaks out of the interior of Croatia in Figure 2(b) before calculating the residual density in Figure 2(c).

\begin{figure}
\includegraphics[height=0.91\textheight]{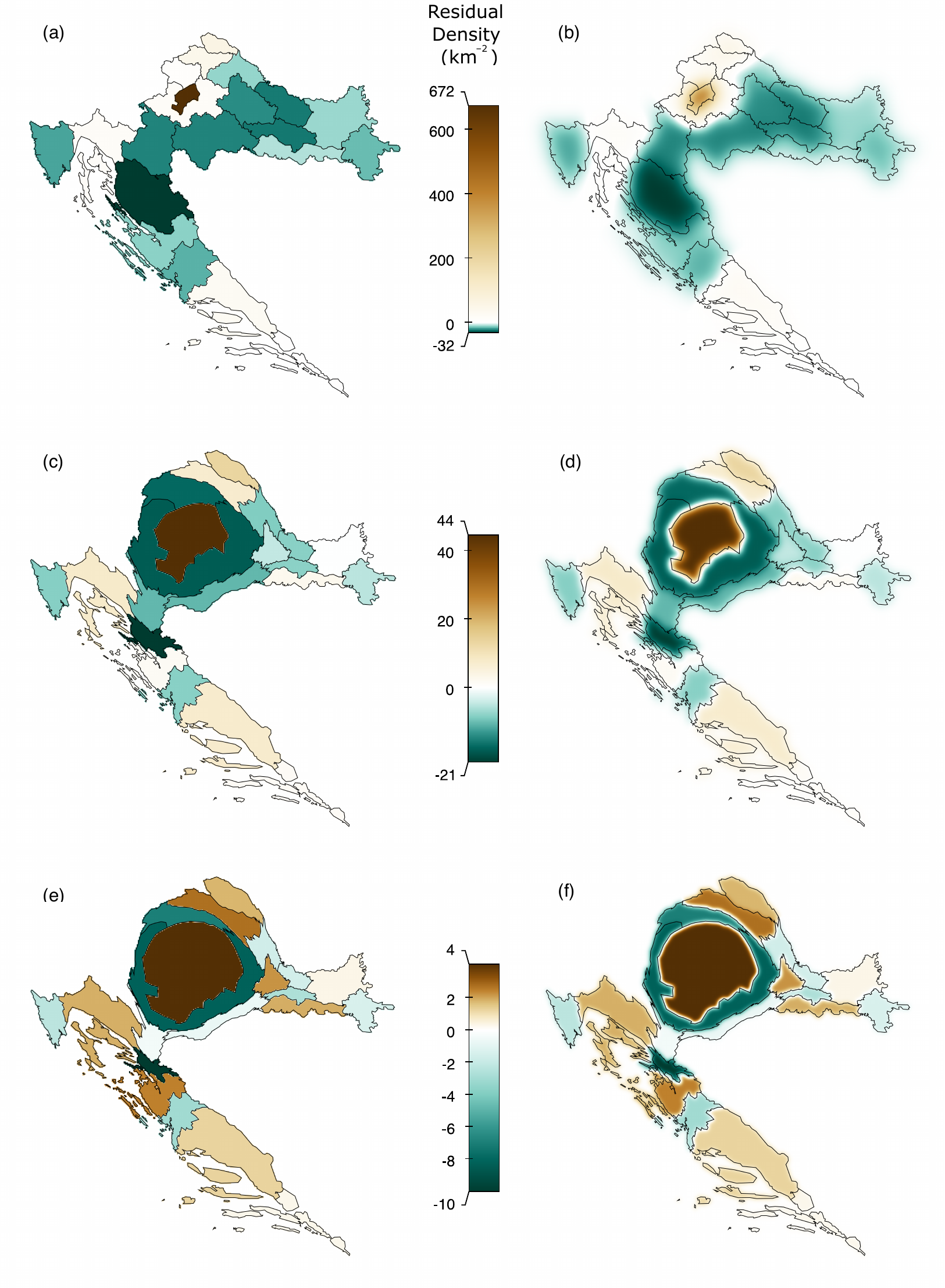}
\caption{Steps in the calculation of a pycnophylactic density of COVID-19 cases in Croatia. Starting from an equal-area map in (a), a contiguous cartogram is iteratively calculated in (b) to (f). Panels (b),(d) and (f) show the residual density after Gaussian smoothing of the densities shown in (a), (c) and (e), respectively.}
\end{figure}

To improve density-equalisation, the computer program by \citeauthor{noauthor_mgastnercartogram-cpp_nodate} (2022) inserts the boundaries obtained from the initial run of the flow-based algorithm as input into a second run; that is, a new piecewise constant residual density \(\rho_\text{res,2}(x, y)\) is constructed; Gaussian smoothing is applied to \(\rho_\text{res,2}(x, y)\), as indicated in Figure 2(d); and a new projection \({\mathbf{t}}_2(x, y)\) is calculated. The newly projected boundaries, shown in Figure 2(e), can be reinserted into the algorithm, which again calculates a piecewise constant residual density \(\rho_\text{res,3}(x, y)\) before applying Gaussian smoothing, as shown in Figure 2(f). With each iteration $k$, discontinuities of \(\rho_{\text{res},k}(x, y)\) tend to become smaller. Thus, the width of the Gaussian kernel that is used for smoothing can be made smaller after each iteration until the width becomes indistinguishable from zero. The colour bars in Figure 2 show that the residual density converges towards zero as the procedure is repeated.

Denoting the projections calculated in the $k$-th iteration of the flow-based algorithm by \({\mathbf{t}}_k(x, y)\), a solution to equation (2) can be obtained by the function composition \(\mathbf{t}(x,y)={\mathbf{t}}_{l}\circ{\mathbf{t}}_{l-1}\circ\ldots\circ{\mathbf{t}}_1(x, y)\), provided that $l$ is sufficiently large. In practice, values of \(l\) between 5 and 20 are usually sufficient to make the areas of all enumeration units on the cartogram accurate to within 1\%. The composed projection \(\mathbf{t}(x,y)\) is differentiable at the boundaries of the enumeration units because each projection \({\mathbf{t}}_k(x, y)\) is the result of Gaussian convolution and, hence, differentiable \parencite{gwosdek_theoretical_2012}. The computer program approximates the Jacobian determinant \(\det{J_{\mathbf{t}}}\) as the factor by which the areas of cells in a fine-grained square grid increase (\(\det{J_{\mathbf{t}}}>1\)) or decrease (\(\det{J_{\mathbf{t}}}<1\)) because of \(\mathbf{t}(x,y)\). In the case of Croatia, the program uses a grid with 512 horizontal and 512 vertical lines. Afterwards, equation (7) is used to solve for \(\rho(x, y)\), which results in a finite-size approximation of a differentiable pycnophylactic density. Figure 1(b) shows the result for COVID-19 cases in Croatia.

\section{Conclusion}

The program by \citeauthor{noauthor_mgastnercartogram-cpp_nodate} (2022) is written in C++ and optimised for computational speed. For the map shown in Figure 1(b), the calculation needed only an average time of 5.1 seconds for a test run on a MacBook Pro laptop with a 2.7 GHz Quad-Core Intel i7 processor. We acknowledge that a comparison with \citeauthor{tobler_smooth_1979}'s (\citeyear{tobler_smooth_1979}b) algorithm would require careful benchmarking of run times, which we still need to implement. However, our preliminary studies suggest that the method described above requires fewer iterations than \citeauthor{tobler_smooth_1979}'s (\citeyear{tobler_smooth_1979}b) algorithm. This observation is explained by the fact that, in the cartogram-based algorithm outlined above, the density associated with any point in space already changes after the first iteration, even for points far from any boundary, because of the significant width of the initial Gaussian kernel. By contrast, in Tobler’s (1979a) algorithm, a grid point that is \(m\) grid spacings away from the nearest boundary requires \(O(m)\) iterations until its density is affected by a neighbouring enumeration unit. This advantage of the cartogram-based algorithm is partly offset by the time needed to calculate Fourier transforms. However, we hypothesise that the substantial reduction in the number of overall iterations more than compensates for the cost of the Fourier transforms. As \cite{tobler_rejoinder_1979} himself noted, ‘The use of a fast Fourier transform $\ldots$ should be investigated to hasten convergence of the algorithm.’

We also acknowledge that pycnophylactic densities generated by the algorithm outlined above do not optimise any predefined criterion for smoothness, whereas \citeauthor{tobler_smooth_1979}'s (\citeyear{tobler_smooth_1979}b) algorithm directly minimises
\[\iint\left[\left(\partial^{2} \rho / \partial x^{2}+\partial^{2} \rho / \partial y^{2}\right)^{2}\right]\, d x\, d y.\]
However, the density \(\rho\) obtained from our algorithm could be used as input to \citeauthor{tobler_smooth_1979}'s (\citeyear{tobler_smooth_1979}b) algorithm, thereby combining the benefits of both methods. The potential advantages of combining the algorithms will be investigated in future work.

\section{Acknowledgements}

This work was supported by the Singapore Ministry of Education (AcRF Tier 1 Grant IG18PRB104, R-607-000-401-114), a Yale-NUS College research award (through grant number A-0000177-00-00), and the Yale-NUS Summer Research Programme. We would like to thank Editage (www.editage.com) for English language editing. We are grateful to Adam Tonks and two anonymous reviewers for their insightful comments and suggestions.

\section{Conflicts of interest}

All authors declare that they have no conflicts of interest.

%\bibliographystyle{abbrvnat}
\printbibliography

\end{document}